\documentclass[a4paper,11pt]{article}
\usepackage{pos}

\usepackage{calrsfs,euscript,mathrsfs}
\usepackage{gensymb}
\usepackage{sectsty}
\usepackage{booktabs}

\title{Exploring the Galactic neutrino flux origins using IceCube datasets}
 \ShortTitle{Exploring the Galactic neutrino flux origins using IceCube datasets}

\author{The IceCube Collaboration \\{\normalsize \normalfont(a complete list of authors can be found at the end of the proceedings)}\\}

\emailAdd{adesai@icecube.wisc.edu}
\abstract{Astrophysical neutrinos detected by the IceCube observatory can be of Galactic or extragalactic origin. The collective contribution of all the detected neutrinos allows us to measure the total diffuse neutrino Galactic and extragalactic signal. In this work, we describe a simulation package that makes use of this diffuse Galactic contribution information to simulate a population of Galactic sources distributed in a manner similar to our own galaxy. This is then compared with the sensitivities reported by different IceCube data samples to estimate the number of sources that IceCube can detect. We provide the results of the simulation that allows us to make statements about the nature of the sources contributing to the IceCube diffuse signal.

\vspace{4mm}
{\bfseries Corresponding authors:}
Abhishek Desai$^{1}$, Jessie Thwaites$^{1*}$, Justin Vandenbroucke$^{1}$\\
{$^{a}$ \itshape Dept. of Physics and Wisconsin IceCube Particle Astrophysics Center, University of Wisconsin{\textendash}Madison, Madison, WI 53706, USA}\\[4mm]
$^*$ Presenter

\ConferenceLogo{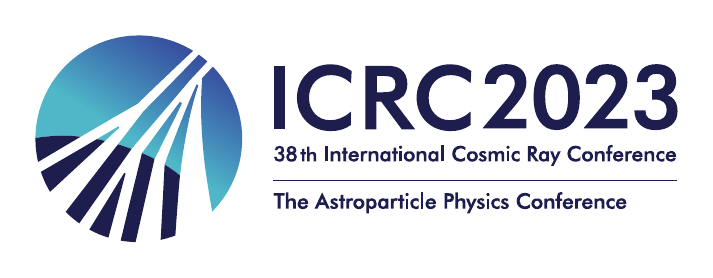}

\FullConference{%
38th International Cosmic Ray Conference (ICRC2023)\\
  26 July - 3 August, 2023\\
  Nagoya, Japan}
}

\begin{document}
\maketitle

\section{Introduction}
\label{sec:intro}

The Milky Way Galaxy is home to numerous objects and matter that can lead to the production of neutrinos  observable on Earth. Recently, \cite{DNN_cascades} reported the observation of neutrino emission from the Galactic plane by the IceCube Neutrino Observatory, with  $>4\sigma$ significance. A confirmation of the exact nature of the sources producing these neutrinos was not found. However, speculations exist about these neutrinos originating from sources like pulsar wind nebula (PWN) or supernova remnants (SNR) or due to cosmic ray interactions (see, for example, \cite{ccsn_Thompson,cosmic_ray_neutrinos_1993,ali_galactic_review}). Fortunately, because of the kiloparsec scale distances of sources and matter distributed across the Galaxy, these speculations can be tested and commented upon (see, for example, \cite{markus_galactic_dnn,snuggy}). In this work, we build upon the results we reported in \cite{snuggy} using additional IceCube data to comment on the nature of these neutrino-producing sources and the future of Galactic neutrino searches using IceCube.

The IceCube neutrino observatory classifies detected neutrino events into track-like or cascade-like depending on the observed event signature \citep{icecube_daq}. Track-like events have better angular resolution due to them being produced by long-lived muons that travel several kilometers in the ice. On the other hand, cascade events are short-lived but have better energy resolution. 
Different IceCube data samples are created using events of one of these types, along with improved reconstructions using methods like cascades with neural network \citep{DNN_cascades} or track events with boosted decision tree \citep{estes_icrc21}. There also exist combined samples like \cite{pierpaolo_combined}, which include both track and cascade events. All these datasets have different effective areas along with different energy and angular resolution, giving them different sensitivities and discovery potentials required to make a detection. In this work, we use the sensitivity and discovery potential curves of 4 different IceCube datasets, namely: 10-year Point Source Tracks ("PST" from this point; see also \cite{icecube_10yr_data}), DNN Cascades ("DNN" from this point; see also \cite{DNN_cascades}), Enhanced Starting Tracks Event Selection ("ESTES" from this point; see also  \cite{estes_icrc19,estes_icrc21}) and a combined event selection made up of ESTES tracks, DNN cascades, and Northern tracks ("Combined sample" from this point; see also  \cite{pierpaolo_combined}). Note that the PS Tracks dataset used here is one reported in \cite{icecube_10yr_data} and cannot be directly compared with the Northern tracks dataset in the combined sample by \cite{pierpaolo_combined}.

\section{Neutrinos from the Galactic center}
\label{sec:back-of-envelope}

Following the procedure shown in \cite{snuggy}, we first estimate simply the number of sources required to make up the observed neutrino emission, provided that all the sources are concentrated at the Galactic center. The observed neutrino emission is taken from the best-fit flux reported by \cite{DNN_cascades} for the KRA$^{50}_\gamma$ template. This is because the reported signal using the KRA$^{50}_\gamma$ template \cite{kra_gamma} is more prominent at the center of the Galaxy. 
The best-fit flux for the KRA$^{50}_\gamma$ template at 100 TeV is given by $\sim$1.5$\times$10$^{-15}$\,TeV$^{-1}$cm$^{-2}$s$^{-1}$ (Fig. 5 of \cite{DNN_cascades}). For each of the tested IceCube datasets, we use the 90\% sensitivity and the 5$\sigma$ discovery potential curves (as reported by \cite{pierpaolo_combined}) to determine the neutrino flux of sources making up the background assuming all sources have equal flux (as all sources are at the center the luminosities are also equal). "Flux" here denotes the differential neutrino number flux in units of TeV$^{-1}$cm$^{-2}$s$^{-1}$ at 100 TeV. As all the sources contribute to the total galactic neutrino flux equally, the number of sources making up the signal is derived by taking the ratio of the total flux and the per-source flux contribution. 

\begin{table}[]
    \centering
    \begin{tabular}{ lcc} 

\toprule
 Sample Tested & $E^{-2.0}$ Flux at $\sim$ 28$\degree$ & $N_{src}$ ($\gamma=-2.0$) \\ 
\midrule
\midrule
 DNN sensitivity & 1.07e-16 & 14 \\
  \midrule
    DNN 5$\sigma$ DP & 4.68e-16 & 3 \\
\midrule
    PST sensitivity & 2.25e-16 & 7  \\
\midrule
    PST 5$\sigma$ DP & 7.97e-16 & 2  \\
\midrule
    ESTES sensitivity & 4.90e-16 & 3 \\
\midrule
    ESTES 5$\sigma$ DP & 1.41e-15 & 1 \\
\midrule
    Combined sensitivity & 8.54e-17 & 18  \\
\midrule
    Combined 5$\sigma$ DP & 4.08e-16 & 4  \\
    \bottomrule
\end{tabular}

    \caption{Assuming that all point sources making up the observed Galactic neutrino signal are concentrated at the center, we show for each dataset the per source flux and the number of sources ($N_{src}$). The flux spectrum ($dN/dE$) is taken to be a power law with an index of 2.0. Note that as no detected Galactic neutrino sources exist, the flux estimates are taken from the sensitivity curve and treated as upper limits, while the number of sources contributing to the signal should be treated as lower limits.
    }
    \label{tab:backofenvelope}
\end{table}

When compared directly to the results presented in \cite{snuggy}, this work includes more information in the form of the ESTES and Combined samples, along with 5$\sigma$ discovery potential curve (as opposed to the 4$\sigma$ discovery potential) for the DNN sample(see \cite{pierpaolo_combined} for more details). Note that in the event that these sources are detected, they will all be clustered at the center leading to source confusion. 

\section{Simulating neutrino sources}
\label{sec:SNuggy}

 We now simulate the neutrino sources in the Galaxy and compare them to sensitivity and discovery potential curves. Following the procedure described in \cite{snuggy}, we use the "\textit{Simulation of the Neutrino and Gamma-ray Galactic Yield}" (\textit{SNuGGY}\footnote{https://github.com/adesai90/SNuGGY}) package. In this work, to simulate source positions, we make use of a modified exponential spatial distribution given by 
\begin{equation}
\rho(R,z) = \rho_0 \left(\frac{R}{R_\odot}\right)^\alpha exp\left(-\beta \frac{R-R_0}{R_\odot}\right) exp\left(-\frac{|z|}{h}\right),
\label{eq:mod_exp_simulation}
\end{equation}
where $R$ and $z$ are the horizontal and vertical scaling lengths respectively, and $\alpha = 2$, $\beta = 3.53$, and $h=0.181$ are parameters for the distribution given by \cite{Lorimer_pulsars, Ahlers_mod_exp}. The Jacobian factor is included while estimating the source positions, causing a shift away from the Galactic center for the $R$ values. 

\begin{figure}[ht!]
   \begin{center}
   \begin{tabular}{c}
    \includegraphics[width=.51\textwidth,trim={1.9cm 0.1cm 0.0cm 1.0cm},clip]{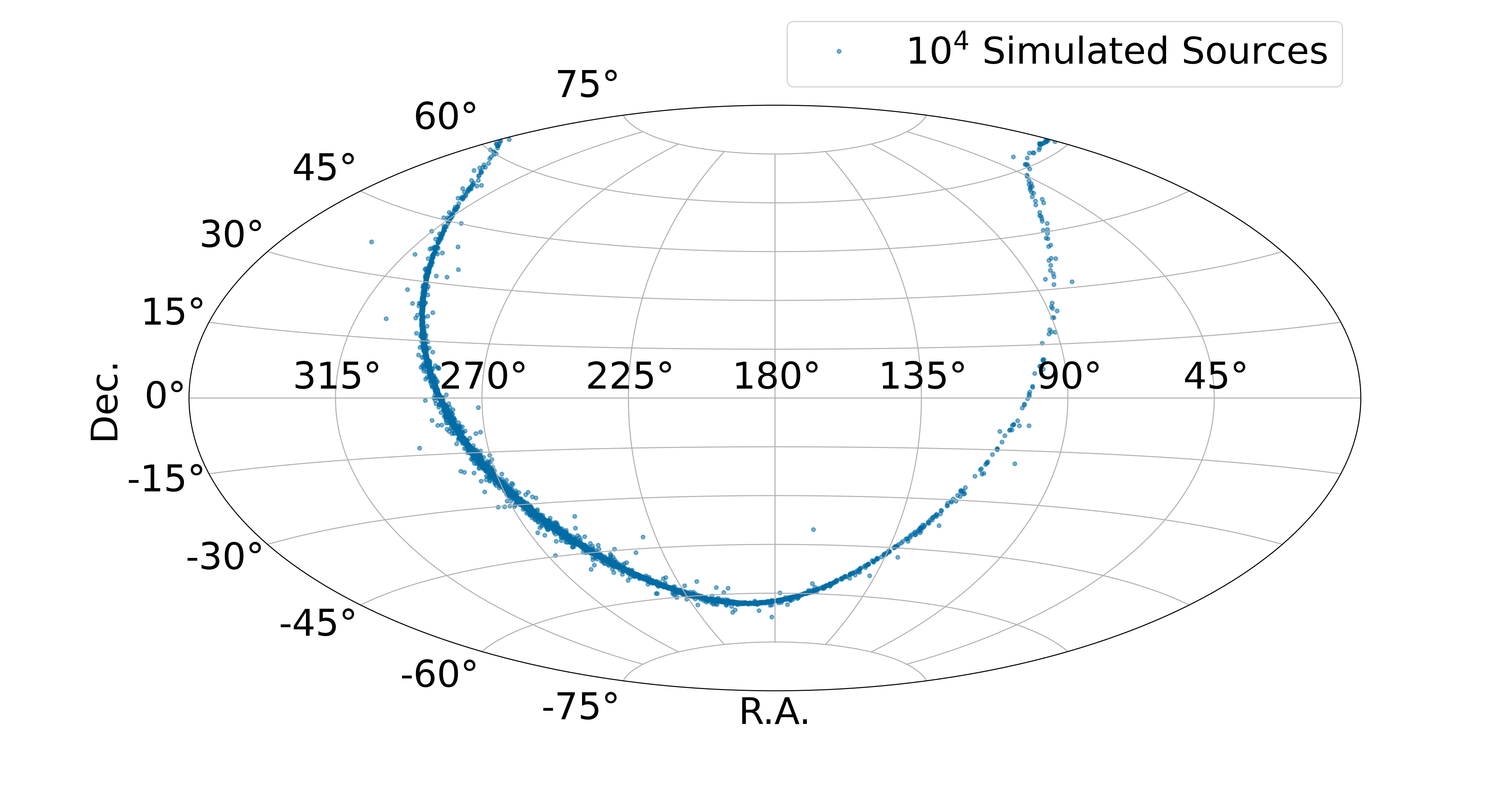}
    \includegraphics[width=.51\textwidth,trim={0.0cm 0.0cm 0.0cm 0.0cm},clip]{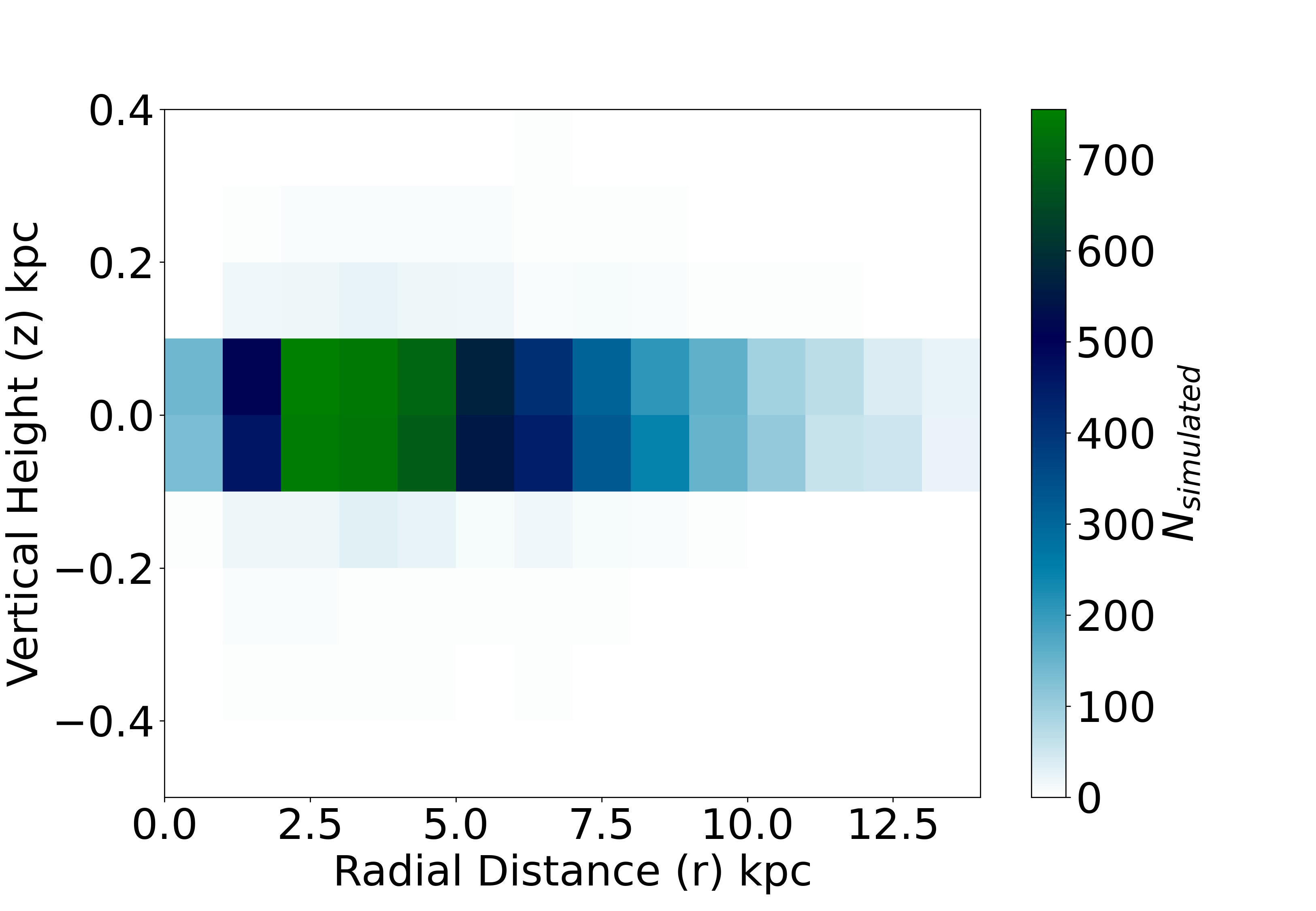}
   \end{tabular}
   \end{center}
   
  \caption{One simulation of $10^4$ sources derived using the SNuGGY framework is shown here. The source coordinates in Galactic coordinates are shown on the left while the distribution of the scale height  $z$ and length $R$ from equation Eq.~\ref{eq:mod_exp_simulation} is shown on the right. Note the shift in the peak away from $R=0$ due to the inclusion of the Jacobian factor.}
  \label{fig:one_simulation_example}
\end{figure}

The neutrino fluxes are simulated using a log-normal luminosity function where the luminosity is defined as the integrated value over an energy range of 10$\,$TeV-10$\,$PeV, and has units of erg/s. The luminosity distribution is given by
\begin{equation}
     P_{LN}(L)=\frac{log_{10}e}{\sigma_L L \sqrt{2\pi}} exp\left(\frac{-(log_{10}L - log_{10}L_{0})^2}{2 \sigma_L^2}\right),
\end{equation}
 where the $L_0$ is the mean luminosity while the $\sigma_L$ parameter controls the width of the distribution. The mean luminosity is calculated using 
 \begin{equation}
    \label{eq:fixed_sc}
    L_{SC}=\frac{\phi_{Galactic}}{\sum\limits_{i=1}^N{\frac{1}{4\pi d_{i}^2}}},
\end{equation}
where $\phi_{Galactic}$ is the total diffuse flux and $N$ is the number of simulated sources at a distance $d_i$. Note that giving a very low value of $\sigma_L$ will reduce the width of the distribution and result in simulated luminosities equal to the mean luminosity (with slight deviations), mimicking a standard candle approach. For more details regarding how the source positions and neutrino fluxes are simulated, see \cite{snuggy}.

\begin{figure}[ht!]
    \centering
    \includegraphics[width=0.88\textwidth,trim={0.0cm 0.2cm 0.0cm 0.2cm},clip]{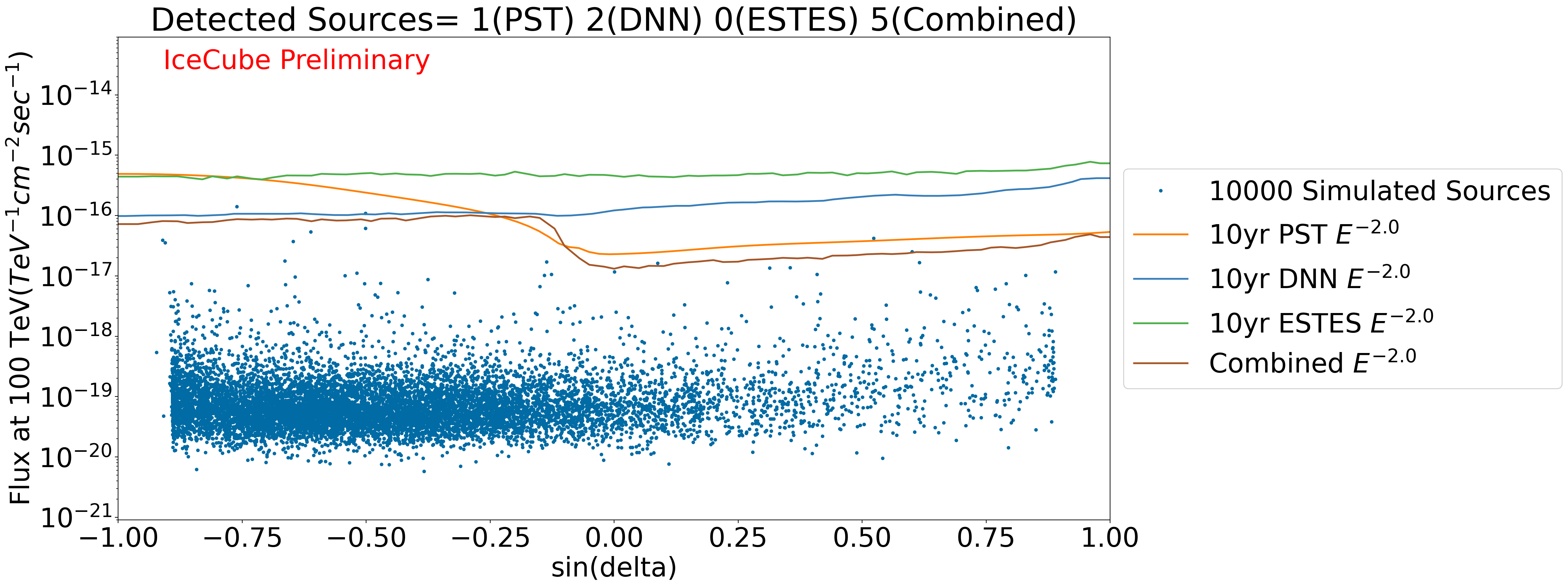}

  \caption{Comparison of simulated sources with the 90\% CL sensitivity curves from the four IceCube data samples. Blue points show $10^4$ simulated sources with fluxes derived using a log-normal luminosity distribution with $\sigma_L$=0.01 (similar to a SC scenario). If a simulated source flux is above the sensitivity curve, the source is counted as detected. The 10yr PST sensitivity is taken from \cite{10yrPStracks_tessa}, DNN from \cite{DNN_cascades} and ESTES and Combined from \cite{pierpaolo_combined}. 
  \label{fig:one_simulation}}
\end{figure}

For each simulated test case, we fix the number of simulated sources. The SNuGGY simulation ensures that the sources have a spatial distribution as shown in Fig.~\ref{fig:one_simulation_example} along with a log-normal luminosity distribution. The simulated differential fluxes at 100 TeV are then used to compare with the sensitivity and discovery potential curves of the IceCube datasets. Two hypotheses are tested here: (1) All sources are close to the center of the Galaxy and (2) sources follow a PWN distribution. The latter will allow us to make assumptions about Galactic neutrino source classes as a whole, as the spatial distribution of galactic sources is similar to each other.

While a simple comparison of the simulated fluxes with the sensitivity or discovery potential fluxes is possible, actual detection of neutrinos from a source is subject to Eddington bias \cite{nora_eddington_bias}. This bias is particularly seen for the "large number of dim sources" case, as it is seen as upward Poisson fluctuations in the number of detected neutrinos. We account for this by estimating the number of energy-integrated neutrino events over a period of 10 years using the simulated neutrino flux and IceCube effective area and adding Poisson fluctuations to the simulated data. The effective area measurements are taken from references \cite{icecube_10yr_data} and \cite{pierpaolo_combined}. If the number of Poisson fluctuated neutrino events are higher than the threshold number of events derived using the sensitivity or discovery potential, the source is considered to be "detected". This calculation is repeated multiple times in the form of a Monte Carlo simulation to derive the mean number of detected sources along with a $1\sigma$ standard deviation.

\begin{figure*}[ht!]
   \begin{center}
   \begin{tabular}{c}
    \includegraphics[width=.49\textwidth,trim={2.2cm 1.3cm 3.5cm 0.0cm},clip]{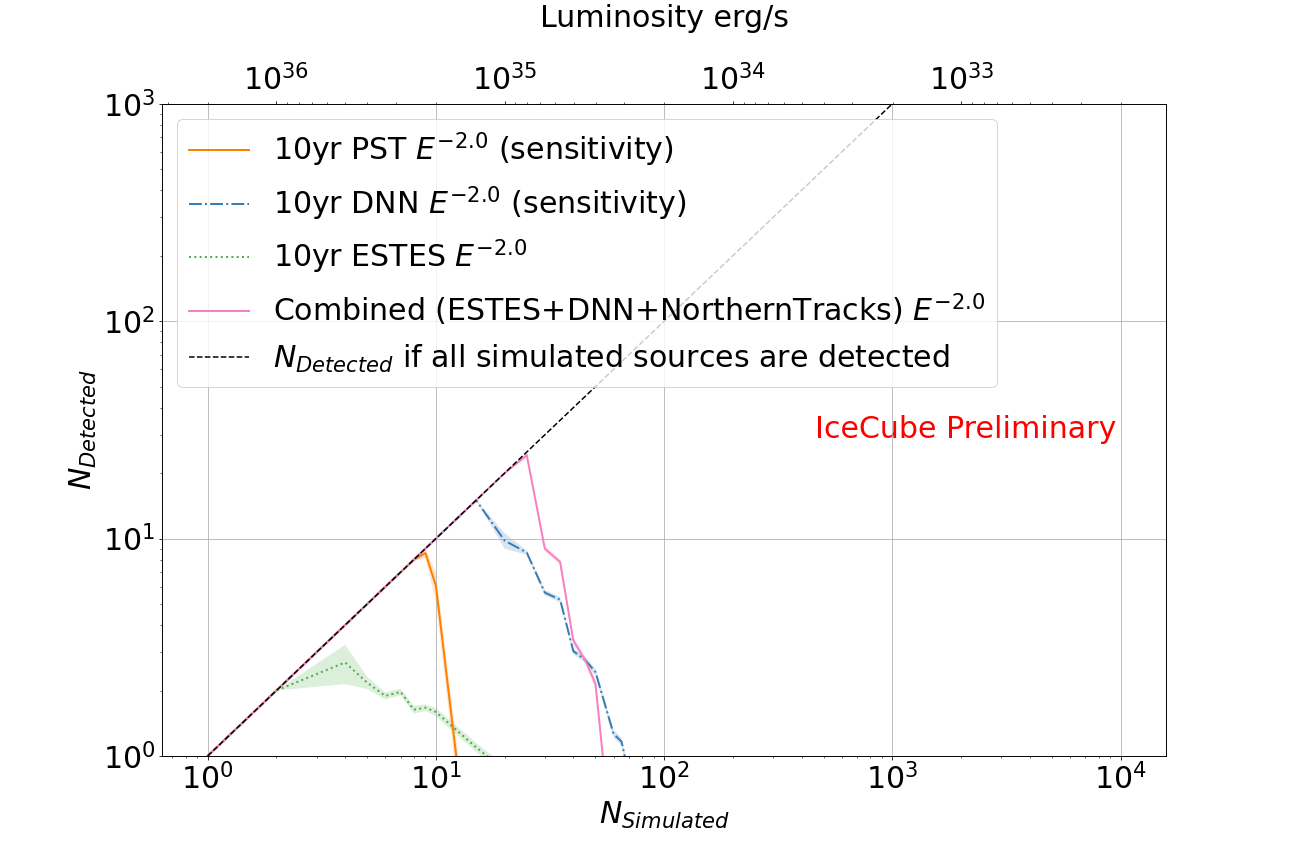} 
    \includegraphics[width=.49\textwidth,trim={2.2cm 1.3cm 3.5cm 0.0cm},clip]{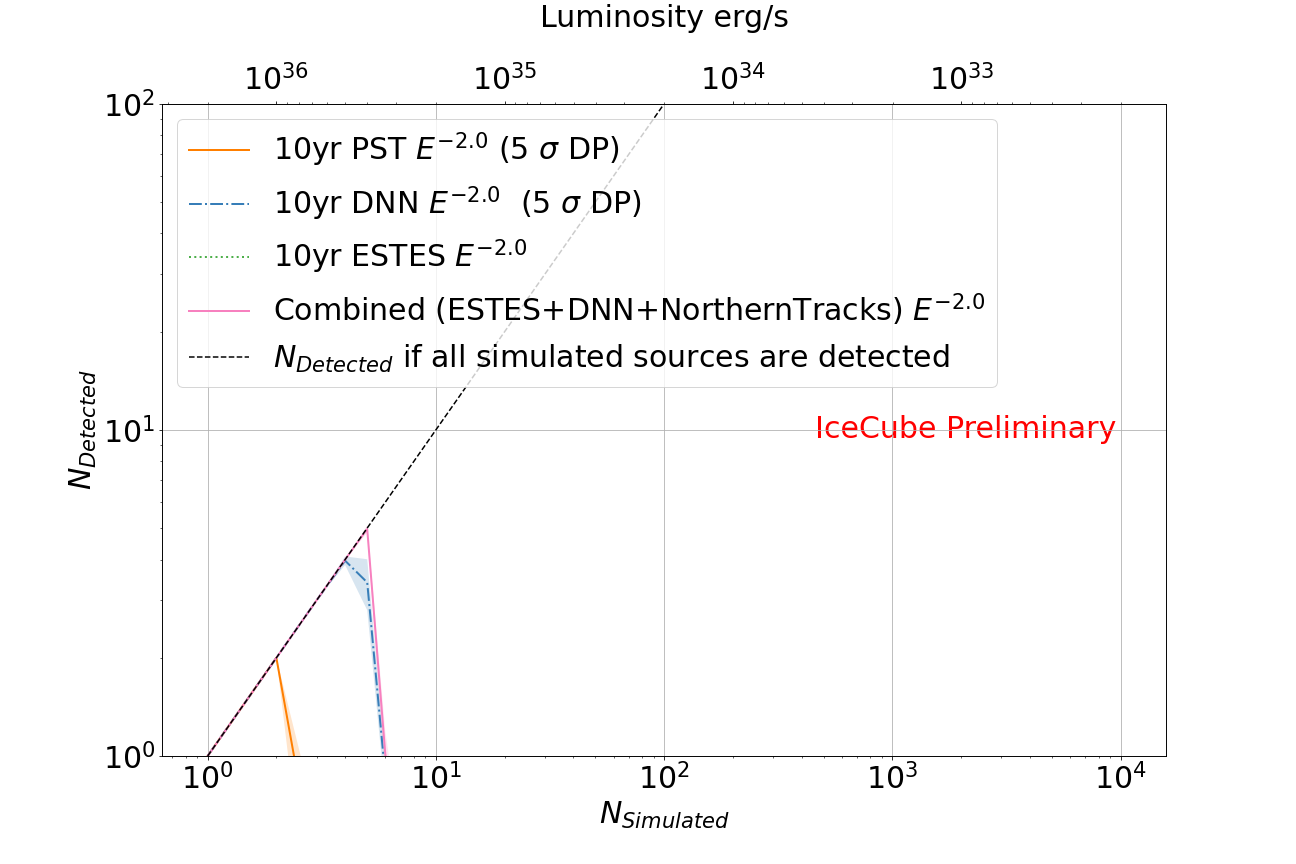}
   \end{tabular}
   \end{center}

  \caption{Special case for sources simulated at the Galactic center:
  The number of detected neutrino sources at the Galactic center for different sensitivity and discovery potential curves (taken from \cite{10yrPStracks_tessa,DNN_cascades, pierpaolo_combined}) while using a $\sigma_L$=0.01\,TeV$^{-1}$cm$^{-2}$s$^{-1}$. The left plot makes use of sensitivity curves, while the right plot makes use of discovery potential curves. 
  The shaded regions show the $\pm 1\sigma$ uncertainty. Note that, in this analysis, the PS Tracks dataset used is taken from \cite{10yrPStracks_tessa} and cannot be directly compared with the Northern Tracks dataset in the combined sample, which uses track-like events in the northern hemisphere with an updated reconstruction. 
  \label{fig:center_case_num_src_detected}}
\end{figure*}

The results for the sources close to the center of the Galaxy are given in Fig.~\ref{fig:center_case_num_src_detected}. The results for this case match the numbers shown in Table.~\ref{tab:backofenvelope}, which is expected. One can see that because of the dependence on the IceCube datasets as a function of declination, the PS tracks sample cannot detect any sources (above $N_{Simulated}\sim3$ for the DP curves) while the other samples can (up to $N_{Simulated}\sim$8). This is because of the better sensitivity of the DNN and Combined datasets in the Southern Hemisphere.

Next, we show the case for a simulation using a $\sigma_L$ of 0.01 and 0.5, which simulates the luminosity distribution as a standard candle or log-normal distribution.

\begin{figure*}[ht!]
   \begin{center}
   \begin{tabular}{c}
    \includegraphics[width=.49\textwidth,trim={2.2cm 1.3cm 3.5cm 0.0cm},clip]{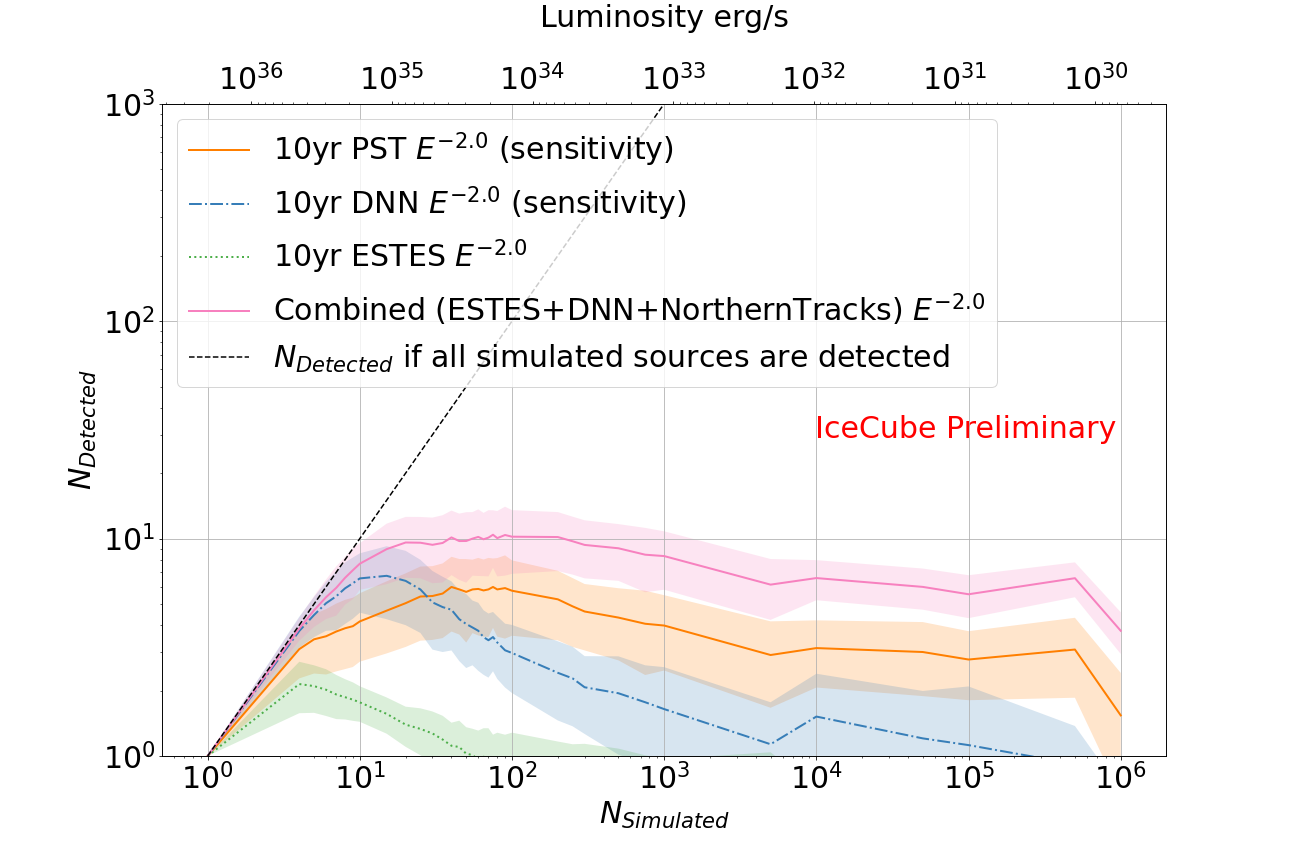} 
    \includegraphics[width=.49\textwidth,trim={2.2cm 1.3cm 3.5cm 0.0cm},clip]{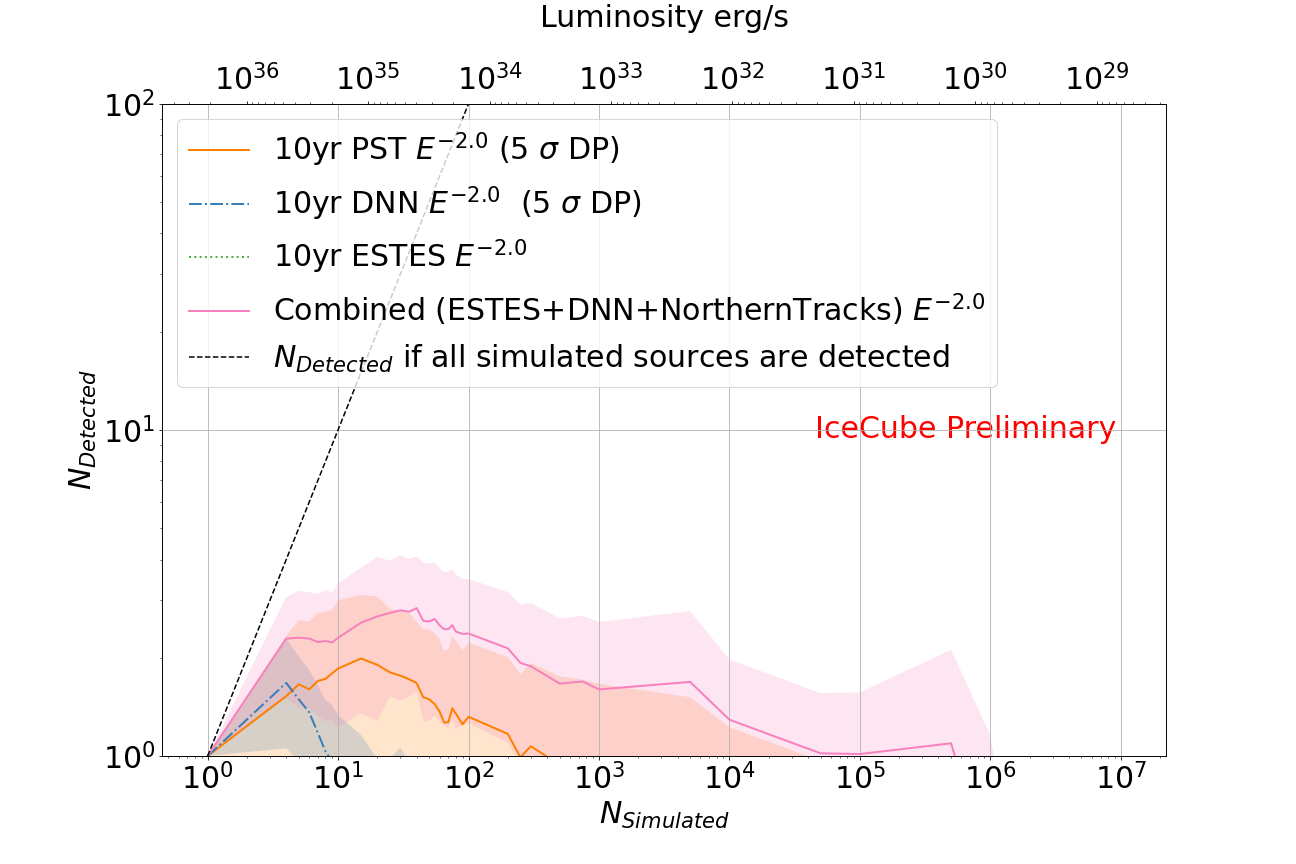}
   \end{tabular}
   \end{center}

  \caption{Case for sources simulated with a realistic geometric distribution and a standard candle approach for fluxes: Number of detected neutrino sources using a $\sigma_L$=0.01 and total diffuse flux equals 2.18$\times10^{-15}$ TeV$^{-1}$cm$^{-2}$s$^{-1}$ is compared with different IceCube sensitivity (left) and discovery potential (right) curves, taken from \cite{10yrPStracks_tessa,DNN_cascades, pierpaolo_combined}. The 2.18$\times10^{-15}$ is obtained using the best-fit neutrino flux derived for the DNN cascade sample using the $\pi^0$ template \cite{DNN_cascades}  
  The shaded regions show the 1$\sigma$ uncertainty.
  \label{fig:sig_0.01_num_detected}}
\end{figure*}

\begin{figure*}[ht!]
   \begin{center}
   \begin{tabular}{c}
    \includegraphics[width=.49\textwidth,trim={2.2cm 1.3cm 3.5cm 0.0cm},clip]{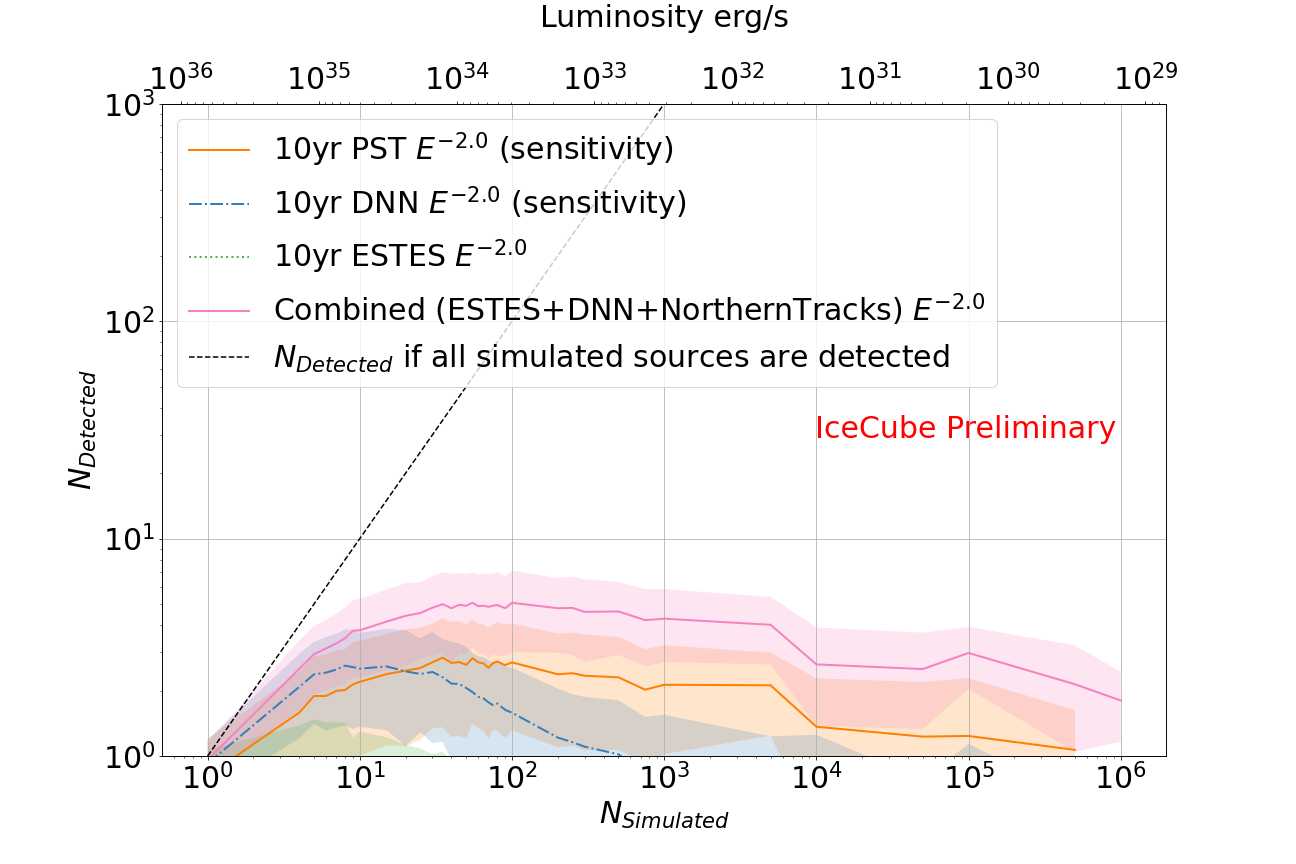} 
    \includegraphics[width=.49\textwidth,trim={2.2cm 1.3cm 3.5cm 0.0cm},clip]{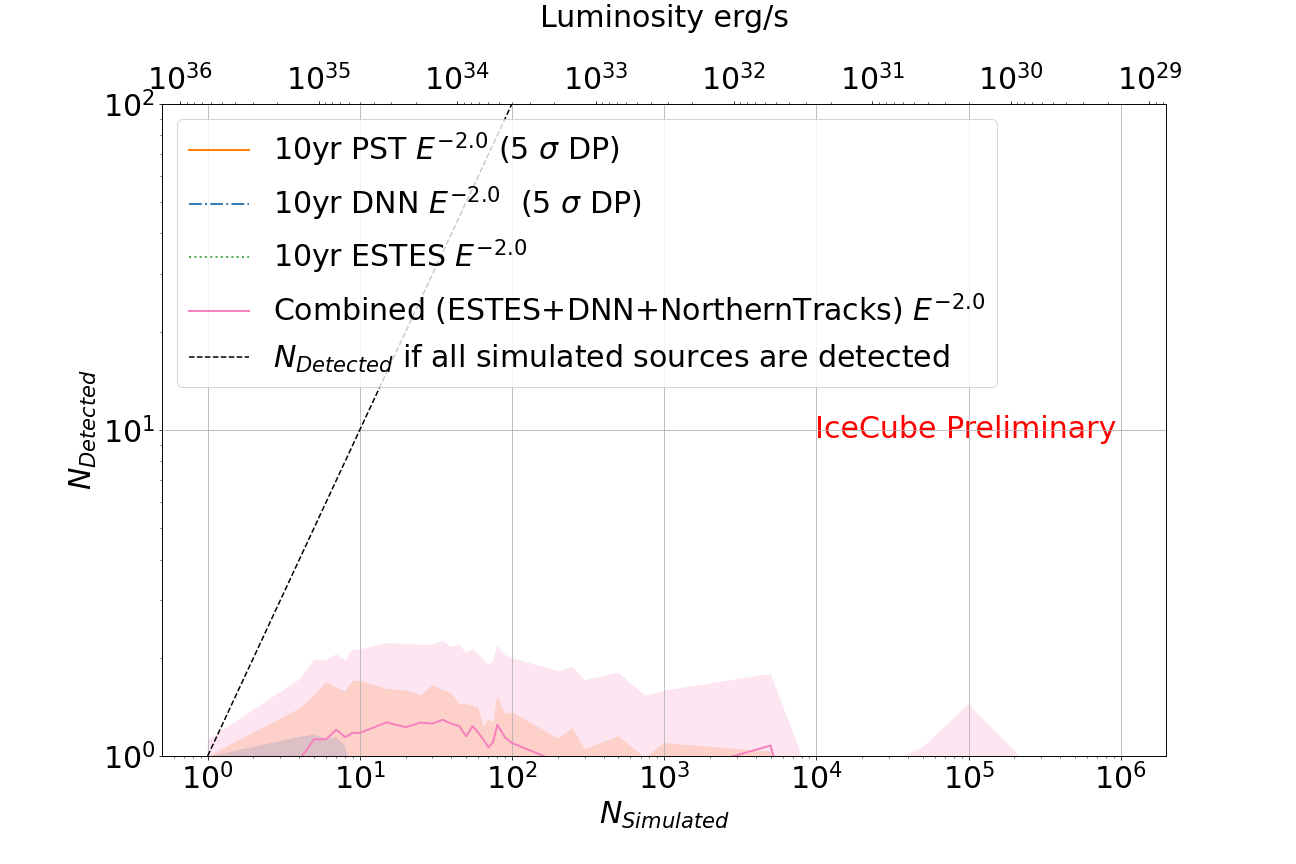}
   \end{tabular}
   \end{center}

  \caption{Case for sources simulated with a realistic geometric distribution and a log-normal approach for fluxes: Number of detected neutrino sources using a $\sigma_L$=0.5 and total diffuse flux equals 2.18$\times10^{-15}$ TeV$^{-1}$cm$^{-2}$s$^{-1}$ is compared with different IceCube sensitivity (left) and discovery potential (right) curves, taken from \cite{10yrPStracks_tessa,DNN_cascades, pierpaolo_combined}.
  The shaded regions show the 1$\sigma$ uncertainty. 
  \label{fig:sig_0.5_num_detected}}
\end{figure*}

\section{Angular Resolution (source confusion)}
\label{sec:angular_res}

The above simulations estimate the number of detected sources after accounting for Eddington bias and do not take into account the different angular resolutions of the IceCube datasets. As shown in \cite{snuggy}, the resolving power of the PS tracks dataset is the best, being able to resolve $\sim26$ sources at 100 TeV close to the center, while ESTES and DNN samples are able to resolve $\sim13$ and $\sim5$ sources respectively.
However, because of the improved sensitivity of the DNN sample in the southern hemisphere, sources at the center of the Galaxy are more likely to be detected by DNN cascades. Finally, as the combined sample is a culmination of DNN cascades, ESTES and Northern Tracks, if a source is detected by the combined sample and DNN or ESTES samples, the combined sample will have an equal or better resolving power due to the increased statistics.

\section{Discussion}
\label{sec:discussion}

In this work, we simulate Galactic source populations to understand the results presented by \cite{DNN_cascades}. While the ESTES and Combined samples are being used in similar studies (to \cite{DNN_cascades}), we can already use our simulations to explore science cases based on different outcomes of the two studies.

Reference \cite{DNN_cascades} was not able to detect any Galactic sources but detected a neutrino signal from the Galactic Plane. Our simulation shows that if $\lesssim5$ sources were making up the total neutrino signal, the DNN cascades sample will be able to detect and resolve them. We could extend this calculation to put an upper limit on the number of sources that can be detected by the dataset. This is done by finding the point where the mean number of detected simulated sources, shown in Figs.~\ref{fig:center_case_num_src_detected}-\ref{fig:sig_0.5_num_detected}, reaches $\sim1$. We report this lower limit in  Table.~\ref{tab:limit_on_src} along with calculations using the ESTES and Combined sample datasets.

\begin{table}[]
    \centering
    \begin{tabular}{llccc}
    \toprule
    Dataset & Quantity & SC   &  LN  & Sources at \\
    & & ($\sigma_L=0.01$) & ($\sigma_L=0.5$) & center \\
     \midrule
     \midrule
    DNN & $N_{src}$ & 8 & 0 & 6\\
    
     & $L_{mean}$ & 1.4$\times10^{35}$ & - & 3.5$\times10^{35}$\\
    \midrule
    ESTES & $N_{src}$ & 1 & 0 & 1\\
    
    & $L_{mean}$ & 2.0$\times10^{36}$ & - & 2$\times10^{36}$\\
    \midrule
    Combined & $N_{src}$ & 1.2$\times10^5$ & 235 & 6\\
    
    &$L_{mean}$ & 6.04$\times10^{30}$ & 1.08$\times10^{33}$ & 3.5$\times10^{35}$\\
    \bottomrule
    \end{tabular}
    \caption{Lower limit on the approximate number of sources (along with the upper limit on the luminosities) detected by each of the datasets based on the simulations. The spectral index used is fixed to 2.0.
    Note that the results shown in \cite{snuggy} use the 4$\sigma$ discovery potential curves while this work makes use of the 5$\sigma$ discovery potential curve for the DNN cascades sample giving a different limit.}
    \label{tab:limit_on_src}
\end{table}

Our simulation shows that the Combined Sample has the best chance of detecting sources. Based on the combined estimates from Table~\ref{tab:backofenvelope} and \ref{tab:limit_on_src}, we find that the increase in the number of detected sources comes mostly from better sensitivity in the Northern Hemisphere (similar to what is seen using the PS tracks sample). The simulations using DP curves shown in Figs.~\ref{fig:center_case_num_src_detected}-\ref{fig:sig_0.5_num_detected} (right) show that the combined sample outperforms PST in the Northern hemisphere too (seen for larger $N_{Simulated}$ values where flux per source is lower). This is because the combined sample makes use of the Northern Tracks IceCube dataset along with the additional data from DNN and ESTES, which improves the sensitivity. However, we already know that \cite{DNN_cascades} and \cite{10yrPStracks_tessa} were not able to detect any sources. Using that fact, if we assume that no sources are detected by the combined sample, we could put a limit of $10^5$ SC sources or a limit of $\sim235$ sources following a LN distribution. We can also put an upper limit on the luminosity of the source population to be of the order of $10^{33}$ erg/s.
This limit is more constraining because of the improvements in the sensitivity of the combined sample due to the combination of the cascades and tracks datasets, which makes it most sensitive in both the Northern and Southern hemispheres.
We can see that from the results depicted here, along with theoretical studies like \cite{Sudoh_galactic_plane_pevetrons,snuggy,markus_galactic_dnn}, future improvements and measurements from IceCube are key to understanding exactly how many sources or points of neutrino emission contribute to the Galactic neutrino flux and their nature.

\bibliographystyle{ICRC}
\setlength{\bibsep}{1ex}
\bibliography{references}

\providecommand{\href}[2]{#2}\begingroup\raggedright\begin{thebibliography}{10}

\bibitem{DNN_cascades}
{\bfseries IceCube} Collaboration, R.~Abbasi {\em et~al.}
  \href{http://dx.doi.org/10.1126/science.adc9818}{{\em Science} {\bfseries
  380} no.~6652, (2023) 1338--1343}.

\bibitem{ccsn_Thompson}
T.~A. Thompson, A.~Burrows, and P.~A. Pinto
  \href{http://dx.doi.org/10.1086/375701}{{\em Astrophys. J.} {\bfseries 592}
  (2003) 434}.

\bibitem{cosmic_ray_neutrinos_1993}
G.~Domokos, B.~Elliott, and S.~Kovesi-Domokos
  \href{http://dx.doi.org/10.1088/0954-3899/19/6/010}{{\em Journal of Physics
  G: Nuclear and Particle Physics} {\bfseries 19} no.~6, (Jun, 1993) 899}.

\bibitem{ali_galactic_review}
A.~Kheirandish \href{http://dx.doi.org/10.1007/s10509-020-03816-3}{{\em
  Astrophys. Space Sci.} {\bfseries 365} no.~6, (2020) 108}.

\bibitem{markus_galactic_dnn}
A.~Ambrosone, K.~M. Groth, E.~Peretti, and M.~Ahlers
  \href{http://dx.doi.org/10.48550/arXiv.2306.17285}{{\em arXiv e-prints}
  (June, 2023) }.

\bibitem{snuggy}
A.~Desai, J.~Vandenbroucke, S.~Anandagoda, J.~Thwaites, and M.~J. Romfoe
  \href{http://dx.doi.org/10.48550/arXiv.2306.17305}{{\em arXiv e-prints}
  (June, 2023) }.

\bibitem{icecube_daq}
{\bfseries IceCube} Collaboration, R.~Abbasi {\em et~al.}
  \href{http://dx.doi.org/10.1016/j.nima.2009.01.001}{{\em Nucl. Instrum. Meth.
  A} {\bfseries 601} (2009) 294--316}.

\bibitem{estes_icrc21}
{\bfseries IceCube} Collaboration, R.~Abbasi {\em et~al.}
  \href{http://dx.doi.org/10.22323/1.395.1130}{{\em PoS} {\bfseries ICRC2021}
  (2021) 1130}.

\bibitem{pierpaolo_combined}
{\bfseries IceCube} Collaboration, P.~Savina {\em et~al.} {\em PoS} {\bfseries
  ICRC2023} (these proceedings) 1010.

\bibitem{icecube_10yr_data}
{\bfseries IceCube} Collaboration, R.~Abbasi {\em et~al.}
  \href{http://dx.doi.org/10.21234/CPKQ-K003}{{\em arXiv e-prints} (Jan., 2021)
  }.

\bibitem{estes_icrc19}
{\bfseries IceCube} Collaboration, M.~Silva and S.~Mancina
  \href{http://dx.doi.org/10.22323/1.358.1010}{{\em PoS} {\bfseries ICRC2019}
  (2020) 1010}.

\bibitem{kra_gamma}
D.~Gaggero, D.~Grasso, A.~Marinelli, A.~Urbano, and M.~Valli
  \href{http://dx.doi.org/10.1088/2041-8205/815/2/L25}{{\em Astrophys. J.
  Lett.} {\bfseries 815} no.~2, (2015) L25}.

\bibitem{Lorimer_pulsars}
D.~R. Lorimer {\em et~al.}
  \href{http://dx.doi.org/10.1111/j.1365-2966.2006.10887.x}{{\em Mon. Not. Roy.
  Astron. Soc.} {\bfseries 372} (2006) 777--800}.

\bibitem{Ahlers_mod_exp}
M.~Ahlers, Y.~Bai, V.~Barger, and R.~Lu
  \href{http://dx.doi.org/10.1103/PhysRevD.93.013009}{{\em Phys. Rev. D}
  {\bfseries 93} no.~1, (2016) 013009}.

\bibitem{10yrPStracks_tessa}
{\bfseries IceCube} Collaboration, M.~G. Aartsen {\em et~al.}
  \href{http://dx.doi.org/10.1103/PhysRevLett.124.051103}{{\em Phys. Rev.
  Lett.} {\bfseries 124} no.~5, (2020) 051103}.

\bibitem{nora_eddington_bias}
N.~L. Strotjohann, M.~Kowalski, and A.~Franckowiak
  \href{http://dx.doi.org/10.1051/0004-6361/201834750}{{\em Astron. Astrophys.}
  {\bfseries 622} (2019) L9}.

\bibitem{Sudoh_galactic_plane_pevetrons}
T.~Sudoh and J.~F. Beacom
  \href{http://dx.doi.org/10.1103/PhysRevD.107.043002}{{\em Phys. Rev. D}
  {\bfseries 107} no.~4, (2023) 043002}.

\end{thebibliography}\endgroup

\clearpage
\section*{Full Author List: IceCube Collaboration}

\scriptsize
\noindent
R. Abbasi$^{17}$,
M. Ackermann$^{63}$,
J. Adams$^{18}$,
S. K. Agarwalla$^{40,\: 64}$,
J. A. Aguilar$^{12}$,
M. Ahlers$^{22}$,
J.M. Alameddine$^{23}$,
N. M. Amin$^{44}$,
K. Andeen$^{42}$,
G. Anton$^{26}$
% ...
\\
\\
% ...
$^{12}$ Universit{\'e} Libre de Bruxelles, Science Faculty CP230, B-1050 Brussels, Belgium \\
% ...
$^{17}$ Department of Physics, Loyola University Chicago, Chicago, IL 60660, USA \\
$^{18}$ Dept. of Physics and Astronomy, University of Canterbury, Private Bag 4800, Christchurch, New Zealand \\
% ...
$^{22}$ Niels Bohr Institute, University of Copenhagen, DK-2100 Copenhagen, Denmark \\
$^{23}$ Dept. of Physics, TU Dortmund University, D-44221 Dortmund, Germany \\
% ...
$^{26}$ Erlangen Centre for Astroparticle Physics, Friedrich-Alexander-Universit{\"a}t Erlangen-N{\"u}rnberg, D-91058 Erlangen, Germany \\
% ...
$^{40}$ Dept. of Physics and Wisconsin IceCube Particle Astrophysics Center, University of Wisconsin{\textendash}Madison, Madison, WI 53706, USA \\
% ...
$^{42}$ Department of Physics, Marquette University, Milwaukee, WI, 53201, USA \\
% ...
$^{44}$ Bartol Research Institute and Dept. of Physics and Astronomy, University of Delaware, Newark, DE 19716, USA \\
% ...
$^{63}$ Deutsches Elektronen-Synchrotron DESY, Platanenallee 6, 15738 Zeuthen, Germany  \\
$^{64}$ Institute of Physics, Sachivalaya Marg, Sainik School Post, Bhubaneswar 751005, India \\
% ...

\subsection*{Acknowledgements}

\noindent
The authors gratefully acknowledge the support from the following agencies and institutions:
USA {\textendash} U.S. National Science Foundation-Office of Polar Programs,
U.S. National Science Foundation-Physics Division,
U.S. National Science Foundation-EPSCoR,
Wisconsin Alumni Research Foundation,
Center for High Throughput Computing (CHTC) at the University of Wisconsin{\textendash}Madison,
Open Science Grid (OSG),
Advanced Cyberinfrastructure Coordination Ecosystem: Services {\&} Support (ACCESS),
Frontera computing project at the Texas Advanced Computing Center,
U.S. Department of Energy-National Energy Research Scientific Computing Center,
Particle astrophysics research computing center at the University of Maryland,
Institute for Cyber-Enabled Research at Michigan State University,
and Astroparticle physics computational facility at Marquette University;
Belgium {\textendash} Funds for Scientific Research (FRS-FNRS and FWO),
FWO Odysseus and Big Science programmes,
and Belgian Federal Science Policy Office (Belspo);
Germany {\textendash} Bundesministerium f{\"u}r Bildung und Forschung (BMBF),
Deutsche Forschungsgemeinschaft (DFG),
Helmholtz Alliance for Astroparticle Physics (HAP),
Initiative and Networking Fund of the Helmholtz Association,
Deutsches Elektronen Synchrotron (DESY),
and High Performance Computing cluster of the RWTH Aachen;
Sweden {\textendash} Swedish Research Council,
Swedish Polar Research Secretariat,
Swedish National Infrastructure for Computing (SNIC),
and Knut and Alice Wallenberg Foundation;
European Union {\textendash} EGI Advanced Computing for research;
Australia {\textendash} Australian Research Council;
Canada {\textendash} Natural Sciences and Engineering Research Council of Canada,
Calcul Qu{\'e}bec, Compute Ontario, Canada Foundation for Innovation, WestGrid, and Compute Canada;
Denmark {\textendash} Villum Fonden, Carlsberg Foundation, and European Commission;
New Zealand {\textendash} Marsden Fund;
Japan {\textendash} Japan Society for Promotion of Science (JSPS)
and Institute for Global Prominent Research (IGPR) of Chiba University;
Korea {\textendash} National Research Foundation of Korea (NRF);
Switzerland {\textendash} Swiss National Science Foundation (SNSF);
United Kingdom {\textendash} Department of Physics, University of Oxford.

\end{document}